\documentclass[12pt]{iopart}
\usepackage{amssymb,latexsym,graphicx}
\begin{document}
\title{Photoluminescence dispersion as a probe of structural inhomogeneity in silica}
\author{Michele D'Amico$^{1, 2, *}$, Fabrizio Messina$^1$, Marco Cannas$^1$, Maurizio Leone$^{1, 2}$ and Roberto Boscaino$^1$}
\address{$^1$ Dipartimento di Scienze Fisiche ed Astronomiche, Universit\'a di Palermo, Via Archirafi 36, I-90123 Palermo, Italy}
\address{$^2$ Istituto di Biofisica, U. O. di Palermo, Consiglio Nazionale delle Ricerche, Via Ugo La Malfa 153, I-90146 Palermo, Italy}
\ead{$^*$damico@fisica.unipa.it}
\begin{abstract}
We report time-resolved photoluminescence spectra of point defects in amorphous silicon dioxide (silica), in particular the decay kinetics of the emission signals of extrinsic Oxygen Deficient Centres of the second type from singlet and directly-excited triplet states are measured and used as a probe of structural inhomogeneity. Luminescence activity in sapphire ($\alpha$-Al$_2$O$_3$) is studied as well and used as a model system to compare the optical properties of defects in silica with those of defects embedded in a crystalline matrix. Only for defects in silica, we observe a variation of the decay lifetimes with emission energy and a time dependence of the first moment of the emission bands. These features are analyzed within a theoretical model with explicit hypothesis about the effect introduced by the disorder of vitreous systems. Separate estimations of the homogenous and inhomogeneous contributions to the measured emission linewidth are obtained: it is found that inhomogeneous effects strongly condition both the triplet and singlet luminescence activities of oxygen deficient centres in silica, although the degree of inhomogeneity of the triplet emission turns out to be lower than that of the singlet emission. Inhomogeneous effects appear to be negligible in sapphire.
\end{abstract}
\pacs{}
\maketitle

\section{Introduction}
Point defects in amorphous silicon dioxide (SiO$_2$) represent a
fundamental technological issue due to the wide range of applications
of this material in current optical and electronic technologies.
Indeed, the formation of defects, typically triggered by exposure to
ultraviolet (UV) or ionizing radiation, compromises the performance
of SiO$_2$ in optical components, optical fibres, and Metal Oxide Semiconductor
transistors \cite{Erice, nalwa}. On the other side, comparing the
properties of color centres hosted by an amorphous and a crystal
matrix is an issue of considerable interest for basic solid state
physics, still leaving several unanswered questions especially
concerning the interplay between homogeneous and inhomogeneous
optical broadening effects \cite{stoneham}. In a crystal each member
of an ensemble of identical defects experiences the same
local environment and its spectroscopical properties
are properties of the single centre. Hence, the optical properties
of the set of defects, such as the homogeneous absorption or
emission linewidth, determined by the electron-phonon coupling \cite{Erice, nalwa}, must be considered as \emph{homogeneous}
properties of the defects. The situation is different in a glass
where, beside the homogeneous features of the single point defect,
different centres are localized in different environments, possibly
featuring a continuous spectrum of geometric configurations. This
statistical distribution of structural parameters can result in a
further broadening of optical bands (\emph{inhomogeneous effect}).
Thus, in an amorphous solid the observed spectroscopic fingerprint
of a set of nominally identical point defects, i.e. belonging to the
same \emph{species}, is due to the convolution of both homogeneous
and inhomogeneous effects. No general recipe is available to
discriminate the extent of the former with respect to the latter
and, as a consequence, the prominence
of inhomogeneous or homogeneous effects in determining the optical
properties of defects in glasses has been a debated question for a
long time \cite{Erice, nalwa, holeburning} .

In previous papers, we reported optical measurements on a
particular kind of point defect in silica, the so-named Oxygen
Deficient Centre of the second type, shortly ODC(II), performed at
several temperatures both by time-resolved and stationary
luminescence techniques. These studies suggested that the
spectroscopic properties of ODCs are significantly conditioned by
inhomogeneous effects \cite{leone1, leone2, cannizzophilos,
cannizzoSn}. Also, ODC(II) have been observed only in the amorphous phase
of SiO$_2$, so being an interesting model system to investigate glassy-specific
inhomogeneous effects. The ODC(II) exists as an intrinsic defect or
in two extrinsic varieties, and its mostly accepted structural model
consists in a twofold coordinated atom (=X$^{\bullet
\bullet}$)\cite{skuja1984, nishikawa1992, skuja1994}, where X can be
either a Si, a Ge or a Sn atom, belonging to the same isoelectronic
group. In particular, the Ge-ODC(II) defects are responsible of an
intense optical activity in Vis-UV range which is currently
associated to the variation of refraction index in the fibre Bragg
gratings after UV writing \cite{nalwa}.

\begin{figure*}[h!]
\centering \includegraphics[width=9 cm]{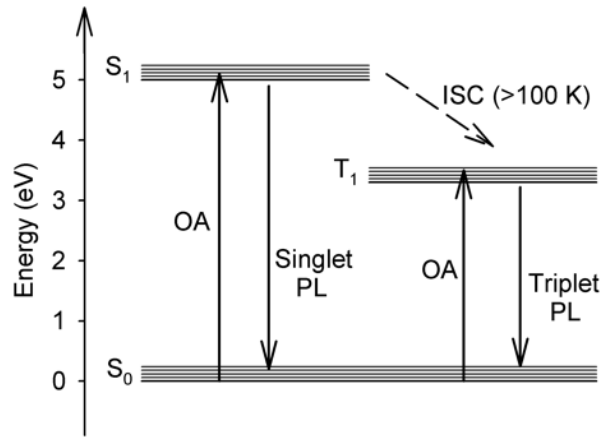}
\caption{Schematic representation of electronic levels related to main optical activities of Ge-ODC(II).}\label{figscheme}
\end{figure*}
In figure \ref{figscheme} are schematically shown the main optical absorption (OA) and photoluminescence (PL) transition for Ge-ODC(II).
A broad nearly-gaussian OA band centred at
$\sim$5.1 eV, assigned to the transition between the ground electronic singlet state
(S$_0$) and the first excited singlet (S$_1$) one, excites a fast
(lifetime in the ns range) PL band centred at $\sim$4.3 eV,
due to the inverse S$_1$$\rightarrow$S$_0$ transition
\cite{skuja1994, SkujaReview98}. Due to partial admixture of S$_1$ and the first
excited triplet state (T$_1$), at T$>$100 K, it is possible also to
populate the T$_1$ state from S$_1$ by a phonon assisted process named
inter-system crossing (ISC). The subsequent radiative decay from
T$_1$ towards S$_0$ gives rise to an additional PL emission centred
at $\sim$3.1 eV.  The lifetime of this band is slow (in the range of
$\mu$s) due to the forbidden spin selection rules for this
T$_1$$\rightarrow$S$_0$ transition \cite{SkujaReview98,
agnello2000}. It is also possible to populate the T${_1}$ state
directly from the ground $S{_0}$ state by exciting with photons of
$\sim$3.7 eV energy. This process has a very low absorption cross
section and bypasses the ISC channel giving rise to the same 3.1 eV
phosphorescence band independently from the temperature of the
system and with the same lifetime  of the
S$_0\rightarrow$S$_1\rightsquigarrow$T$_1\rightarrow$S$_0$ process \cite{SkujaReview98}.

In a recent paper \cite{damicoPRB} we introduced a new experimental
investigation approach, based on time-resolved luminescence
measurements, which was able to yield an estimation of
the homogeneous and inhomogeneous linewidth of the fast (ns
lifetime) ODC(II) luminescence band due to decay from the $S_1$ state. In this paper we generalize our
analysis applying it to the same model defect, i.e. the
ODC(II) in silica, measuring also the "slow" band assigned to the de-excitation from the $T_1$ state. This phosphorescence band is excited directly populating the first triplet
electronic state to avoid possible inhomogeneous effects arising from ISC process \cite{leone2, agnello2003}. Our aim is to find out whether our approach is
applicable also to a slow ($\mu$s lifetime) triplet emission band
and if the extent of inhomogeneous effects affecting triplet and
singlet emission processes are comparable or not. Finally, in order
to compare the results with those obtained in a system where
inhomogeneous effects should be absent, we report the same study
performed on the PL of point defects in crystalline sapphire. Although the luminescence activity of defects in
irradiated or doped sapphire was extensively studied in the past, several aspects about the decay kinetics, defects interconversion processes, band attributions and structural models of the emitting defects are not clear yet \cite{chen, caulfield, surdo}.
\section{Experimental Section}\label{EXP}
We report measurements performed on a fused silica sample
(commercial name: Infrasil 301, $5\times5\times1$ mm sized and
provided by Heraeus Quartzglas \cite{heraeus}). This sample, hereafter named
I301, is manufactured by fusion and quenching of natural quartz, with
typical concentration of impurities $\sim$20 ppm in weight. In
particular, as-grown I301 features a $\sim$1 ppm concentration of Ge
impurities, due to contamination of the quartz from which the
material was produced. Previous studies demonstrated that a consistent portion of
the Ge impurities are arranged as Ge-ODC(II) defects in the as-grown
material \cite{SkujaReview98, sgjncs03}. For comparison with defects
in a crystal we used also an as-grown commercial sapphire sample
($\alpha$-Al$_2$O$_3$) provided by A.D. Mackay INC (Broadway, New
York) and rod flame polished \cite{mackay}.
\\Photoluminescence measurements were done in a standard back-scattering geometry, under excitation by a pulsed laser (Vibrant OPOTEK: pulsewidth of 5 ns, repetition rate of 10 Hz) tunable in the UV-Visible range. The luminescence emitted by the samples was dispersed by a spectrograph (SpectraPro 2300i, PI Acton, 300 mm focal length) equipped by three different gratings, and detected by an air-cooled intensified CCD (Charge-Coupled Device PIMAX, PI Acton). The detection system can be electronically gated so as to acquire the emitted signal only in a given temporal window defined by its width ($t_W$) and by its delay $t$ from the laser pulse.
\\The luminescence from the triplet excited state of Ge-ODC(II) in the I301 sample was collected under laser excitation (energy density per pulse of 1.00$\pm$0.02 mJ/cm$^2$) at 330 nm (3.75 eV), corresponding to the S$_0$$\rightarrow$T$_1$ absorption peak, and using a 300 grooves/mm grating (blaze at 500 nm) with a 4 nm spectral bandwidth. The PL decay was followed by performing different acquisitions with the same integration time $t_W$=15 $\mu$s but at different delays $t$, going from 0 to 300 $\mu$s from the laser pulse.
\\Analogous measurements on the luminescence arising from the singlet excited state of Ge-ODC(II) was collected under laser excitation (energy density per pulse of 0.30$\pm$0.02 mJ/cm$^2$) at 240 nm (5.17 eV), corresponding to the S$_0$$\rightarrow$S$_1$ absorption peak, and using the same 300 grooves/mm grating (blaze at 500 nm) with a 3 nm spectral bandwidth. In this
case different spectra were acquired with $t_W$=1 ns and $t$ going from 0 to 60 ns.
\\The luminescence of the sapphire sample was acquired using
a 230 nm (5.40 eV) excitation wavelength (energy density per pulse of 1.00$\pm$0.02 mJ/cm$^2$), and a 150 grooves/mm
grating (blaze at 300 nm) with a 8 nm spectral bandwidth. The decay was followed varying $t$ from 0 to 400 $\mu$s and with $t_W$=4 $\mu$s.
All the spectra were corrected for spectrograph dispersion and for instrumental response. All measurements reported here were performed on samples kept at 25 K in high vacuum ($\sim10^{-6}$ mbar) within a helium continuous flow cryostat (Optistat CF-V, OXFORD Instruments).

\section{Results}
\begin{figure*}[h!]
\centering \includegraphics[width=14 cm]{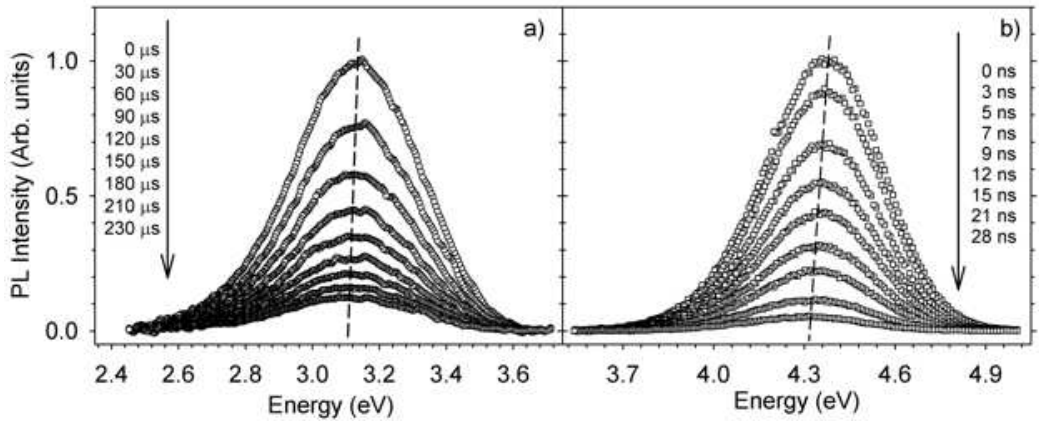} \caption{Time
evolution of the lineshape of triplet (a) and singlet (b) Ge-ODC(II)
luminescence signals at 25 K excited at 3.75 eV and 5.17 eV,
respectively. Different spectra detected at different time delays
from the laser pulse are shown. The dashed line follows the position
of PL peaks as a guide to the eye.}\label{figdecay}
\end{figure*}

We show in figure \ref{figdecay}-(a) the time-resolved spectra of the triplet (T$_1$$\rightarrow$S$_0$) PL activity of Ge-ODC(II) in the I301 silica sample.
In figure \ref{figdecay}-(b) analogous measurements performed on the singlet (S$_1$$\rightarrow$S$_0$) PL emission of the same defects are shown. The dashed lines drawn in figure \ref{figdecay} follow the emission peaks at different time delay $t$ and we observe a non-verticality of their slopes, indicating an experimental detectable red shift of both luminescence bands during the decay.

\begin{figure*}[h!]
\centering \includegraphics[width=14 cm]{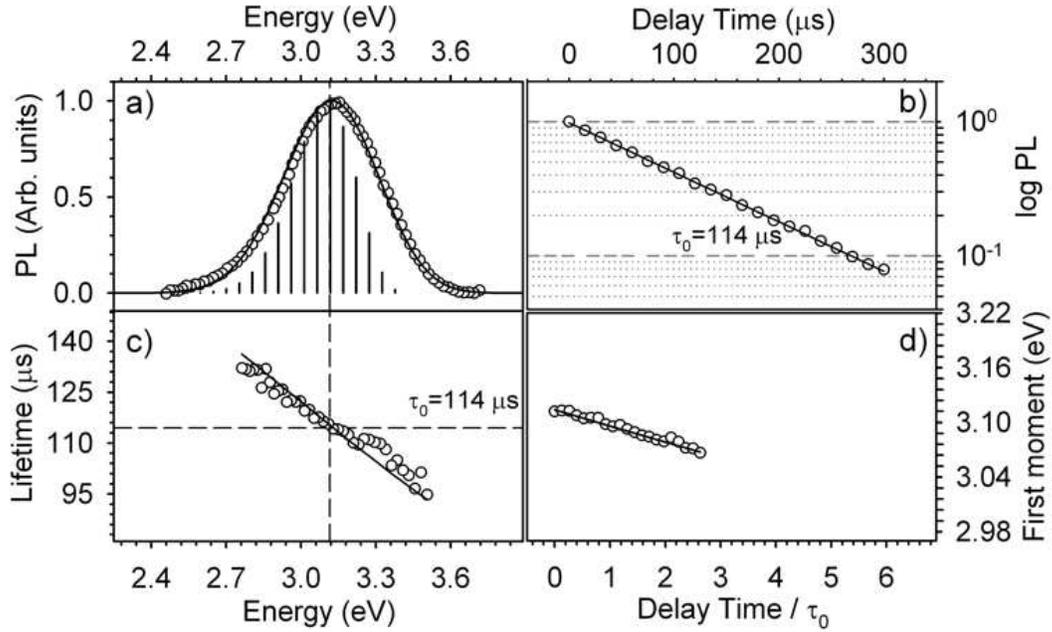} \caption{Panel (a): Luminescence emission lineshape of triplet Ge-ODC(II) activity as measured at T=25 K upon excitation at 3.75 eV, immediately after the end of laser pulse ($t$=0). Panel (b): decay kinetics observed at the peak emission energy ($\sim$3.1 eV). Panel (c): Decay lifetimes as a function of the emission energy. Panel (d): First moment of the emission band as a function of time delay. The continuous lines represent the results of the fitting procedure by our theoretical model (see discussion).} \label{figtriplet}
\end{figure*}

\begin{figure*}[h!]
\centering \includegraphics[width=14 cm]{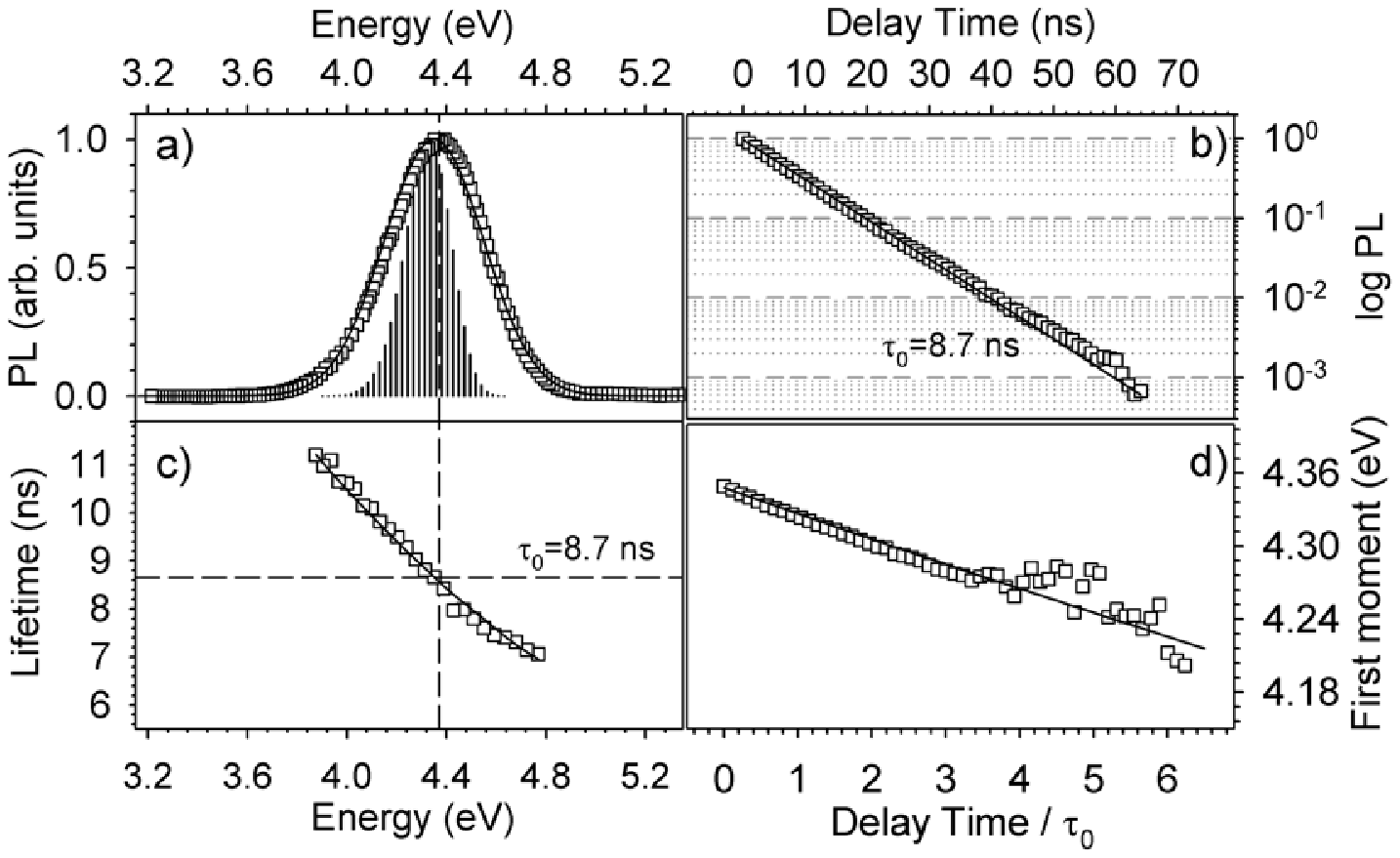}
\caption{Panel (a): Luminescence emission lineshape of singlet Ge-ODC(II) activity as measured at T=25 K upon excitation at 5.17 eV, immediately after the end of laser pulse ($t$=0). Panel (b): decay kinetics observed at the peak emission energy ($\sim$4.4 eV). Panel (c): Decay lifetimes as a function of the emission energy. Panel (d): First moment of the emission band as a function of time delay. The continuous lines represent the results of the fitting procedure by our theoretical model (see discussion).}\label{figsinglet}
\end{figure*}

In figures \ref{figtriplet}-(a) and
\ref{figsinglet}-(a) we report the signal acquired at t=0 for the two PL activities of the Ge-ODC(II), corresponding to the most intense spectra in figures \ref{figdecay}-(a) and \ref{figdecay}-(b),
respectively. The triplet PL band of Ge-ODC(II), as acquired
immediately after the end of the laser pulse, is peaked at $\sim$3.1
eV and features a 0.44 eV width (Full Width at Half Maximum, FWHM),
while the singlet band features a $\sim$4.4 eV peak position and a
0.45 eV FWHM.

\begin{figure*}[h!]
\centering \includegraphics[width=14 cm]{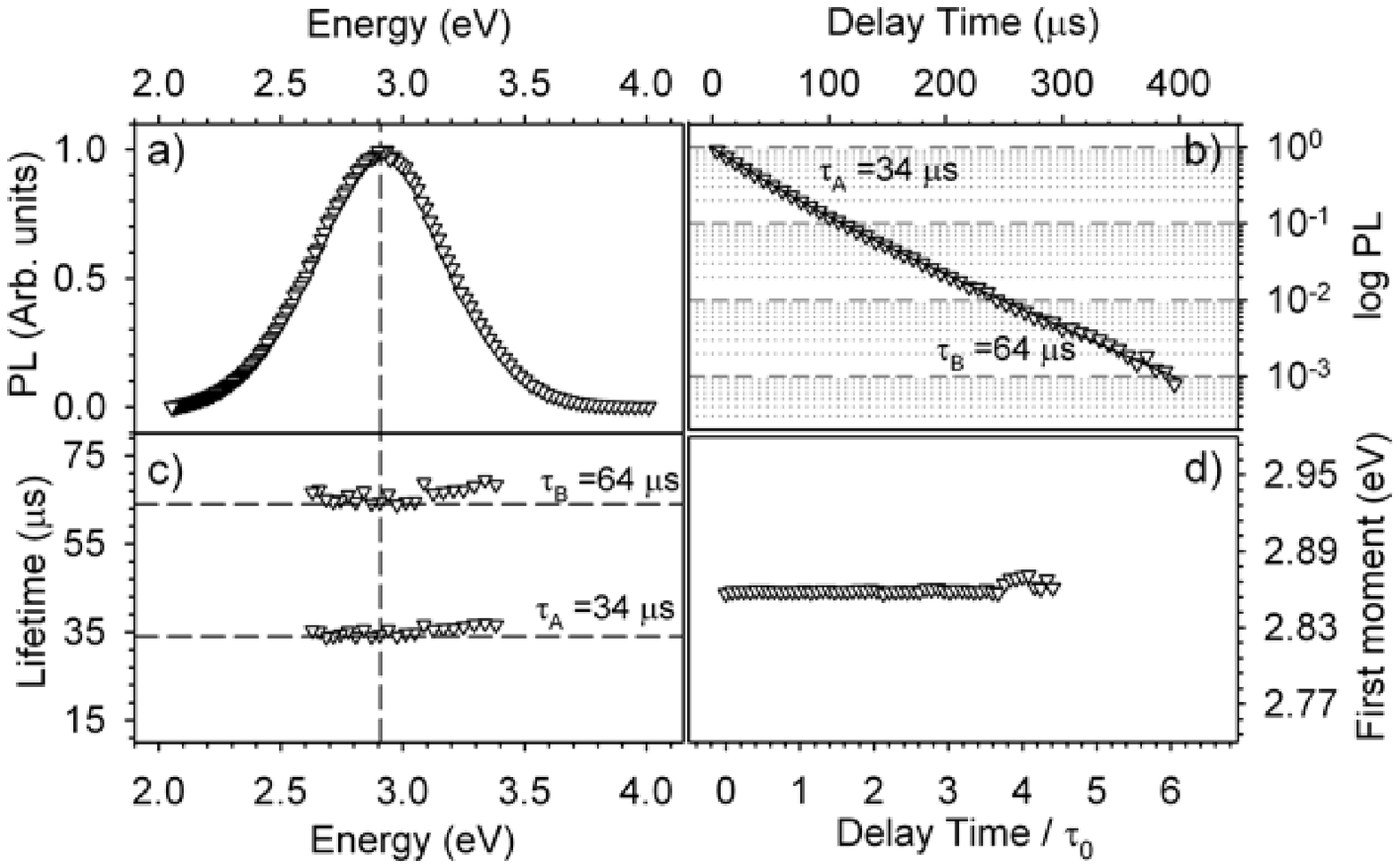} \caption{Panel (a):
Emission lineshape of sapphire PL activity as measured at T=25 K
upon excitation at 5.40 eV, immediately after the end of laser pulse
($t$=0). Panel (b): decay kinetics observed at the peak emission
energy ($\sim$2.8 eV). Panel (c): Decay lifetimes as a function of the
emission energy. Panel (d): First moment of the emission band as a
function of time delay.}\label{figsapphire}
\end{figure*}

Analogous time-resolved measurements were carried out on the PL activity in the sapphire sample with purposes of comparison, and we report in figure \ref{figsapphire}-(a) the spectrum acquired for $t$=0: the PL band observed in sapphire, as acquired immediately after
the end of the laser pulse, is peaked at $\sim$2.9 eV and features a
0.61 eV FWHM. The
spectroscopic parameters of the signal in figure \ref{figsapphire}-(a)
are consistent with a luminescence signal previously observed in
literature, and associated either to the so-called P-centre (an
anion-cation vacancy pair featuring a charge transfer transition)
as proposed by a few works \cite{pujats, springis}, or to an
extrinsic defect as proposed in \cite{evans2}. However, it is
worth noting that the detailed structural model of the centre
responsible for the observed luminescence is not relevant here.
Indeed, to the purposes of the present work we are going to discuss
this signal only as a model of a slow
luminescence of a defect in a crystalline oxide and, as we will see, it presents a $\sim$$\mu$s lifetime.

From all time-resolved measurements one can extract the lifetime
dispersion curves, namely the dependence of the decay lifetime from
emission energy. To this purpose the lifetimes were obtained by a fitting procedure of PL data at a given emission energy, carried out
with a single exponential function for both activities of
Ge-ODC(II), and with a double exponential function for sapphire PL
activity. Representative decays measured at the band peak energies,
3.1 eV, 4.4 eV and 2.8 eV for triplet Ge-ODC(II), singlet Ge-ODC(II)
and sapphire PL signals, are reported with relative fitting curves in figures \ref{figtriplet}-(b),
\ref{figsinglet}-(b), and \ref{figsapphire}-(b) respectively. The
experimental low temperature (25 K) was chosen to ensure purely
radiative decays from excited electronic state, especially to prevent the activation of ISC
process from the excited singlet ODC(II) state (see figure \ref{figscheme}).

The lifetime of the triplet Ge-ODC(II) (figure \ref{figtriplet}-(c)) varies from $\sim$130 to $\sim$100 $\mu$s for emission energies increasing from 2.8 to 3.5 eV, while that of the singlet Ge-ODC(II) (figure \ref{figsinglet}-(c)) varies from $\sim$11 to $\sim$7 ns in the 3.8$\div$4.8 eV range. Finally, in the sapphire sample lifetime dispersion is not observed: indeed, the two lifetimes characterizing the luminescence decay are independent from emission energy: they remain fixed to $\tau_A$=34 $\mu$s and $\tau_B$=64 $\mu$s all over the range of the PL band.\footnote{It is worth noting that the double exponential behavior of P-centre luminescence in sapphire is unknown at the best of our knowledge. In \cite{pujats} the authors found a lifetime of $\sim$50 $\mu$s which is consistent with the mean of the two lifetimes found here. It is beyond the aim of this paper to investigate about the reason behind these decay features.}

The observed energy dependence of the luminescence lifetime is
expected to cause a progressive red shift of the first moment of the
band, due to different temporal evolutions of different parts of the
PL band. From measured spectra we have thus calculated the time
dependence of the first moment of the luminescence bands: the
results are reported in figures \ref{figtriplet}-(d),
\ref{figsinglet}-(d) and \ref{figsapphire}-(d) for triplet
Ge-ODC(II), singlet Ge-ODC(II) and sapphire activities, respectively.
The horizontal axes represent the time delay from the laser pulse in
unit of the lifetime $\tau_0$: for both Ge-ODC(II) activities
$\tau_0$ is defined as the lifetime of PL signal at the central
emission energies, 114 $\mu$s and 8.7 ns for triplet and singlet
decay respectively. For sapphire activity $\tau_0$ is chosen to be
the mean value of the two experimental lifetimes $\tau_A$ and
$\tau_B$. We observe that both slow and fast PL activities in silica
feature an approximately linear red shift of the band as a function
of time, whereas the first moment of the PL activity in sapphire has
a constant value consistent with
the results found for the lifetimes.

\section{Discussion}\label{RES}
The comparison between results on the oxygen deficient centres in
SiO$_2$ and the defects in sapphire suggests that the distribution of lifetimes measured for different
emission energies and the correspondent red-shift of
first moment of the band as a function of delay time, are peculiar features of defects embedded in
amorphous solids as opposed to defects in crystals, where such
effects are not observed. This characteristic behavior of "amorphous
defects" can be conveniently referred to as \emph{luminescence
dispersion}. A similar result had been pointed out in a previous
work \cite{damicoPRB} by comparing the PL properties of the
S$_1$$\rightarrow$S$_0$ transition of Ge-ODC(II) with those of
F-type centres in Lithium Fluoride. While featuring comparable fast
(ns) lifetimes, the decay properties of these two systems turned out
to be very different: a strong luminescence dispersion was found for
Ge-ODC(II) (now confirmed by data in figure \ref{figsinglet}), while
the decay lifetime of F-centres in LiF was found to be independent
from emission energy and the luminescence peak independent from time
\cite{damicoPRB}.

Present results yield a strong generalization of previous findings:
indeed, data reported in section \ref{EXP} demonstrate that the PL
dispersion effect occurs both for the fast (ns) singlet and for the
slow ($\mu$s) triplet luminescence of the Ge-ODC(II), so being
independent from the temporal range of the decay kinetics as well as
from the nature of the transition. Also, dispersion effects are
absent in sapphire defects, similarly to what previously observed in
LiF \cite{damicoPRB}, and notwithstanding the double-exponential
decay kinetics, which may suggest the coexistence of two slightly
different varieties of the optically active centre.

On the whole, our results suggest luminescence dispersion to be a
general optical property which allows to clearly discriminate the
behavior of defects embedded in amorphous solids from crystalline
ones. At least, this appears to be true for oxides. Also, it is
worth stressing that the lifetime of a PL band of defects in a solid
is widely regarded as a strong fingerprint of the defect, which can
be used to unambiguously recognize it by time-resolved PL
measurements. It is worth noting that, without taking care of
lifetime dispersion effects evidenced here, lifetime differences as
large as $\sim$30\% in amorphous systems can wrongly be regarded as
signatures of different defects.

These experimental results can be discussed in the framework of a
theoretical model \cite{damicoPRB} based on the standard background
for description of point defects' optical activities in a solid
matrix \cite{Erice, stoneham}, adding the hypothesis of gaussian
distribution (centre $\widehat{E_0}$ and  half-width $\sigma_{in}$)
of the zero phonon energy $E_0$, that is the energy difference
between ground and first excited electronic state both in the ground
vibrational sub-level. This hypothesis accounts for the effect of the
different environments which can accommodate different members of an
ensemble of point defects in an amorphous matrix. The peak PL
emission energy $E_e$ can be equivalently used as a statistically
distributed parameter instead of the zero phonon energy. Indeed,
these two spectroscopic features are linked by the relation
$E_e=E_0-S$ where $S$ is the half Stokes shift, that is the half
difference between the peak of absorption and luminescence bands and
which is supposed undistributed as the other homogeneous parameters
(the homogeneous half-width $\sigma_{ho}$ and the oscillator
strength $f$). Hence, the gaussian distribution of $E_0$ implies a gaussian distribution of $E_e$ with the same half-width
$\sigma_{in}$ and centred at $\widehat{E_e}=\widehat{E_0}-S$.

On this basis, one can write a quantitative expression of the PL emitted by the ensemble of color centres in an amorphous
solid \cite{damicoPRB}:
\begin{eqnarray}
 L_s(E,t|\widehat{E_e},\sigma_{in},\sigma_{ho},f)\propto \nonumber \\
 \int \left[E^3 P(E|E_e,\sigma_{ho}) \cdot e^{-t/\tau(E_e,\sigma_{ho},f)}\right] \cdot
e^{-\frac{\left(E_e-\widehat{E_e}\right)^2}{2\sigma_{in}^2}}dE_e.
\label{main}
\end{eqnarray}
Equation \ref{main} gives the emission signal $L_s(E,t)$ measured at time $t$ at the spectral position $E$, as obtained by convolution of a gaussian distribution of $E_e$, whose half-width $\sigma_{in}$ represents the inhomogeneous linewidth of the ensemble of defects, with a homogeneous term (within squared parentheses) representing the PL of a defect emitting at a given value of $E_e$ and with homogeneous half-width $\sigma_{ho}$: the spectral shape of the homogeneous term is $E^3 P(E|E_{em},\sigma_{ho})$, where the $P$ function is a Poissonian with first moment $E_{e}$ and second moment $\sigma_{ho}$. Finally, the radiative lifetime is given by the Forster's equation \cite{damicoPRB, forster}:
\begin{equation}
 1/\tau(E_e,\sigma_{ho},f)=\frac{2e^2n^2}{m_ec^3\hbar^2}f \int P(E|E_e,\sigma_{ho})E^3 dE \label{ratetau}
\end{equation}
where $f$ is the oscillator strength of the defect, and $n$ is the refractive index of silica.\footnote{In writing   \ref{ratetau} we have neglected the "effective field correction" term, which can be argued to be close to unity in silica within the investigated spectral range. We also neglect the slight energy dependence of the refraction index $n$, which we assume for silica to be 1.5.} One can numerically integrate \ref{main} in order to simulate the time-resolved PL spectra, $L_s(E,t)$, as a function of the four parameters $\widehat{E_e},\sigma_{in},\sigma_{ho},f$. From $L_s(E,t)$, the decay lifetime $\tau_s(E)$ and the kinetics M$_{1s}$($t$) of the first moment can be easily calculated, by using the same procedure applied to the experimental data $L(E,t)$.

Equation \ref{ratetau} implies that defects with different peak emission
positions $E_e$ within the gaussian distribution should decay with different lifetimes. In this sense, it immediately allows to understand data in figures \ref{figtriplet}-(c) and (d) on a qualitative basis, if we suppose the PL band of ODC(II) as arising from the inhomogeneous overlap of bands peaked at different energies, statistically distributed within the defect population.

The behavior shown in figures \ref{figtriplet} and  \ref{figsinglet}
for silica PL activity can be examined in the frame of the above theoretical model. To
this aim we have performed numerical integration varying the above
mentioned parameters, to obtain a set of three theoretical curves
which simultaneously fit the shape of PL bands, the dependence of the decay lifetime as a function
of the emission energy and the kinetics of the
first moment. This procedure was performed both for the
triplet and singlet luminescence signals of Ge-ODC(II). The
continuous lines in figures \ref{figtriplet} and \ref{figsinglet}
represent the results of our fitting procedure, while the histograms
show the discrete Poissonian homogeneous lineshapes of half width
$\sigma_{ho}$ as obtained by our fit procedure. Since the PL
dispersion effect found here for Ge-ODC(II) is not evidenced in
sapphire luminescence, we argue that in a crystal, beside a few
imperfections due to dislocations or strain, the inhomogeneous
effects (and thus the related width) are virtually absent. As a
consequence the red shift of the first moment of PL band and the
dispersion of lifetimes are not possible, consistently with
experimental results (see figure \ref{figsapphire}).

\begin{table}
\caption{\label{results}Upper section: best fitting parameters obtained by our theoretical model for the investigated PL activities. Lower section: Values of $\lambda$, $\sigma_{tot}$, $\hbar \omega_p$ and $H$, as calculated from best fitting parameters}
\begin{indented}
\item[]\begin{tabular}{@{}lllll}
\br
& $\widehat{E_e}\ [eV]$  & $\sigma_{in}\ [meV]$  & $\sigma_{ho}\ [meV]$  &   f\\
\mr
Triplet PL  & 3.08$\pm$0.05  &  140$\pm$8  & 125$\pm$16  &  (3.3$\pm$0.5) 10$^{-5}$  \\
Singlet PL  & 4.32$\pm$0.05  & 177$\pm$10  &  93$\pm$12  &  0.073$\pm$0.010 \\
\hline\hline
& $\lambda (\%)$ &   $\sigma_{tot} \ [meV]$  & $\hbar \omega_p\ [meV]$  & $H$ \\
\mr
Triplet PL & 56$\pm$4  &  188$\pm$9 &  52$\pm$14 & 6$\pm$2\\
Singlet PL & 78$\pm$5  &  200$\pm$10 &  23$\pm$6 & 17$\pm$5\\
\br
\end{tabular}
\end{indented}
\end{table}
Upper part of table \ref{results} resumes the best parameters
obtained by our fitting procedure for the two luminescence
transitions of Ge-ODC(II). In the lower part of table \ref{results}
we also report the parameter $\lambda=\sigma_{in}^2/\sigma_{tot}^2$
which estimates the degree of inhomogeneity. On one hand, the
results on the parameter $\lambda$ evidence that inhomogeneous
effects strongly affect both the electronic transitions of the
ODC(II) defects in silica. On the other side, the value of $\lambda$
for triplet emission of Ge-ODC(II) (56\%) is smaller than that found for
singlet emission (78\%). Hence, the width of the inhomogeneous
distribution turns out to be greater for the S$_1 \rightarrow$S$_0$
transition than for the T$_1 \rightarrow$S$_0$ one. This finding can
be qualitatively visualized by comparing figure \ref{figtriplet}-(c)
with figure \ref{figsinglet}-(c): in fact, the relative lifetime
increase observed by moving leftwards by a FWHM on the horizontal
axis is lower ($\sim$20\%) for the triplet PL than ($\sim$33\%) for
the singlet PL band.

This result leads to an important consideration about the meaning of
inhomogeneity: the inhomogeneous width has to be considered as a
property of a specific electronic transition occurring at the defect
site, rather than a property of the defect. As a matter of fact, the
physical property of the defect which lies at the root of
inhomogeneity effects is the site-to-site distribution of the
structural parameters, such as bond angles and lengths. In this
sense, the statistical distribution of $E_e$ should be regarded as a
convenient, and synthetic, representation of inhomogeneity effects;
the form and width of such a distribution are determined in
principle by the detailed dependence of the emission peak $E_e$ from the
microscopical structural parameters. Only quantum mechanical
calculations can investigate the form of this mapping function, and
may allow to understand why it ultimately results in a larger degree
of inhomogeneity affecting the S$_1$$\rightarrow$S$_0$ transition as
compared to the T$_1$$\rightarrow$S$_0$ transition of Ge-ODC(II).
Also, more experimental investigations are needed to find out
whether this difference between triplet
and singlet emissions is a general property of defects in amorphous
systems or a peculiar feature of oxygen deficient centres in silica.

Finally, the values of the oscillator strength found here are in
excellent agreement with those reported in a review paper about
oxygen deficiency centres in silica \cite{SkujaReview98}: 0.03-0.07
for the singlet band of Ge-ODC(II) and 1.2$\cdot 10^{-5}$ for the triplet band.
From data in the upper part of table \ref{results} we can also
calculate the Huang-Rhys factor $H=S^2/\sigma_{ho}^2$, the
vibrational frequency $\hbar \omega_p =\sigma_{ho}^2/S$, the total
width (from $\sigma_{tot}^2=\sigma_{in}^2+\sigma_{ho}^2$)
\cite{damicoPRB}. All these quantities are summarized in the lower
part of table \ref{results} and are obtained by using $S$=0.30 eV and 0.38 eV for
the half Stokes shift of triplet and singlet PL band, respectively. The values of $S$ were estimated experimentally by measuring the
half-difference between the spectral positions of the excitation
energies and emission peaks. Result on the vibrational frequencies
shows that the Ge-ODC(II) couple with very low frequency modes,
accordingly with previous experimental and computational results
\cite{cannizzoSn, damicoPRB, galeener, umari}.

\section{Conclusions}\label{CONCL}
We have studied by time-resolved luminescence the extrinsic
(Ge-related) oxygen deficient centres in amorphous silicon dioxide.
Both the triplet and the singlet PL of the defect feature a
dispersion of decay lifetimes within the emission band and a
temporal red shift of their first moments. Comparison with a defect in
crystalline sapphire demonstrates that these effects are peculiar of
a centre embedded in a disordered solid. Experimental findings are
analyzed within a theoretical frame which models the effects induced
by glassy disorder on the optical properties of defects in silica to numerical estimate the homogeneous and inhomogeneous half width. We find that the degree of inhomogeneity
experienced by the triplet luminescence band is appreciably less
than by the singlet one.
\ack
We acknowledge financial support received from project "P.O.R.
Regione Sicilia - Misura 3.15 - Sottoazione C". We are grateful to LAMP research group (http://www.fisica.unipa.it/amorphous/) for support and enlightening discussions. The authors would like to thank G. Lapis and G. Napoli for
assistance in cryogenic work. 
\\
\section*{References}


\begin{thebibliography}{99}
\bibitem{Erice} \emph{Defects in SiO$_2$ and Related Dielectrics: Science and Technology} 2000 eds Pacchioni G,  Skuja L, and Griscom D L (Norwell USA: Kluwer Academic Publishers) 
\bibitem{nalwa}   \emph{Silicon-based Materials and Devices} 2001 ed Nalwa H S (San Diego USA: Academic Press)
\bibitem{stoneham} Stoneham A M \emph{Theory of Defects in Solids} 1975 vol 1 (Oxford: Clarendon Press)
\bibitem{holeburning} \emph{Persistent Spectral Hole-Burning: Science and Applications} (\emph{Topics in current Physics vol 44}) 1988 ed Moerner W E, (Berlin heidelberg: Springer-Verlag)
\bibitem{leone1}Leone M, Agnello S, Boscaino R, Cannas M  and Gelardi F M 1999 \emph{Phys. Rev. B} \textbf{60} 11475
\bibitem{leone2} Cannizzo A, Agnello S, Boscaino R, Cannas M,  Gelardi F M, Grandi S and Leone M 2003 \emph{J. of Phys. and Chem. of Solids} \textbf{64} 2437
\bibitem{cannizzophilos} Cannizzo A and Leone M 2004 \emph{Phil. Mag.} \textbf{84} 1651
\bibitem{cannizzoSn}  Cannizzo A, Leone M, Boscaino R, Paleari A, Chiodini N, Grandi S and Mustarelli P 2006 \emph{J. of Non-Crys. Solids} \textbf{352} 2082
\bibitem{skuja1984} Skuja L N, Streletsky A N and Pakovich A B 1984 \emph{Solid State Commun.} \textbf{50} 1069
\bibitem{nishikawa1992} Nishikawa H, Shiroyama T, Nakamura R, Ohki Y, Nagasawa K and Hama Y 1992 \emph{Phys. Rev. B} \textbf{45} 586
\bibitem{skuja1994} Skuja L 1994 \emph{J. Non-Cryst. Solids} \textbf{179}  51
\bibitem{SkujaReview98} Skuja L 1998 \emph{J. Non-Cryst. Solids} \textbf{239} 16
\bibitem{agnello2000}Agnello S, Boscaino R, Cannas M, Gelardi F M and Leone M 2000 \emph{Phys. Rev. B} \textbf{61} 1946
\bibitem{damicoPRB}D'Amico M, Messina F, Cannas M, Leone M and Boscaino R 2008 \emph{Phys. Rev. B} \textbf{78} 014203
\bibitem{agnello2003} Agnello S, Boscaino R, Cannas M, Gelardi F M, Leone M and Boizot B 2003 \emph{Phys. Rev. B} \textbf{67} 033202
\bibitem{chen} Chen W, Tang H, Shi C,Deng J, Shi J, Zhou Y, Xia S, Wang Y and Yin S 1995 \emph{Appl. Phys. Lett.} \textbf{67} 317
\bibitem{caulfield}Caulfield K J, Cooper R and Boas J F 1997 \emph{J. Phys: Condens. Matter} \textbf{9} 6457
\bibitem{surdo} Surdo A I, Kortov V S, Pustarov V A and Yakovlev V Y 2005 \emph{Phys. Sta. Sol. c} \textbf{2} 527
\bibitem{heraeus} Heraeus Quartzglas, Hanau, Germany, catalog POL-0/102/E
\bibitem{sgjncs03} Grandi S, Mustarelli P, Agnello S, Cannas M, and Cannizzo A 2003 \emph{J. Sol-Gel Sci. Technol.} \textbf{26} 915
\bibitem{mackay} A D Mackay, Inc 7509 North Broadway PO Box 'G' Red Hook, New York 12571-0046
\bibitem{pujats}Pujats A V, Springis M J and Valbis J A 1980 \emph{Phys. Stat. Sol. a} \textbf{62} K85
\bibitem{springis}Springis M, Kulis P, Veispals A and Tale I 1995 \emph{Rad. Measurements} \textbf{24} 453
\bibitem{evans2} BD Evans \emph{J Appl Phys} \textbf{70} 1991, 3995
\bibitem{forster}F$\ddot{o}$rster Th 1951 \emph{Fluoreszenz Organischer Verbindungen} Vandenhoeck und Ruprecht, G$\ddot{o}$ttinen 158
\bibitem{galeener} Galeener F L, Leadbetter A J and Stringfellow M W 1983 \emph{Phys. Rev. B} \textbf{27} 1052
\bibitem{umari} Umari P, Gonze X and Pasquarello A 2003 \emph{Phys. Rev. Lett.} \textbf{90} 027401
\end{thebibliography}
\end{document}